# Dimensional crossover of polaron dynamics in Nb:$SrTiO_3$/$SrTiO_3$ superlattices: Possible mechanism of thermopower enhancement

Woo Seok Choi[1], Hiromichi Ohta[2], Soon Jae Moon[1,a)], Yun Sang Lee[3], and Tae Won Noh[1,*]

[1] *ReCOE, Department of Physics and Astronomy, Seoul National University, Seoul 151-747, Korea*

[2] *Graduate School of Engineering, Nagoya University, Furo-cho, Chikusa, Nagoya 464-8603, Japan and PRESTO, Japan Science and Technology Agency, Kawaguchi 332-0012, Japan*

[3] *Department of Physics, Soongsil University, Seoul 156-743, Korea*

Using optical spectroscopy, we investigated the electrodynamic properties of Nb:$SrTiO_3$/$SrTiO_3$ superlattices. In these superlattices, a large enhancement of the Seebeck coefficient ($S$) has been reported with decreasing Nb:$SrTiO_3$ layer thickness [refer to H. Ohta *et al.*, Nature Mater. **6**, 129 (2007)]. By analyzing the optical spectra, we found that the polaron plays an important role in determining the electrodynamic properties of the superlattices. With decreasing Nb:$SrTiO_3$ layer thickness from eleven to one unit cell, we observed a threefold enhancement of the polaron effective mass and relaxation time. Such increases were attributed to a dimensional crossover of polaron from 3D to quasi-2D. Moreover, the modified nature of the polaron at low dimensions enhanced the thermoelectric properties of the oxide superlattice, by simultaneously increasing $S$ and preventing the decrease of carrier mobility. Our results indicate that strong electron-phonon coupling can provide an alternative pathway in searching efficient thermoelectric materials.

## I. INTRODUCTION

Thermoelectrics has received renewed interest as a candidate for solving the world's demand for sustainable energy.[1-3] In a thermoelectric material, a temperature (*T*) gradient is employed to diffuse the charge carriers from the warmer end to the cooler end of the sample. The consequent accumulation of charges results in an electric field, leading to a voltage difference between the two ends of the sample. Through this process, we can transfer waste heat into electricity without damaging the environment. The figure of merit (*zT*) of a thermoelectric material can be written as

$$zT = \frac{S^2 \sigma_{\mathrm{dc}} T}{\kappa} \qquad (1),$$

where $S$, $\sigma_{\mathrm{dc}}$, and $\kappa$ are the Seebeck coefficient, the *dc* electrical conductivity, and the thermal conductivity, respectively. Note that $S$ is an important physical parameter for obtaining high $zT$ values.

Numerous approaches have been used to search for materials with large $S$ values. It should be noted that $S$ is linearly proportional to effective mass ($m^*$) for metals and degenerate semiconductors.[3, 4] Therefore, many studies have adopted an approach to tune the density of states (DOS) to achieve larger $m^*$, which corresponds to the energy derivative of the DOS near the Fermi energy ($E_F$). In particular, using a semiconductor quantum-well superlattice (SL) or quantum wire, enhanced $S$ values can be obtained by modifying the DOS in low dimensional structures.[5, 6] Such efforts have been recently extended to nanostructured composites.[7] As an alternative approach, transition metal oxides have attracted substantial attention. Large values of $S$ have been reported in doped $Na_xCo_2O_4$ (Ref. 8, 9), where strong electron correlation produce large values of $m^*$. However, both of these conventional approaches have problems in that the increase in $m^*$ is usually accompanied by a decrease in mobility $\mu$ ($\equiv e\tau / m^*$, where $e$ and $\tau$ are the charge and the relaxation time of the free carrier). As indicated in Eq. (1), this in turn leads to a decrease in $\sigma_{\mathrm{dc}}$ ($\equiv ne\mu = ne\tau / m^*$, where $n$ is the carrier density), which makes it difficult to enhance $zT$ (Ref. 3).

Recently, efforts to find efficient thermoelectric materials have also been extended to artificial oxide SLs. One example is an SL composed of conducting Nb:$SrTiO_3$ (20% Nb-doped $SrTiO_3$, Nb:STO) layers and insulating $SrTiO_3$ (STO) layers.[10, 11] In these Nb:STO/STO SLs, $|S|$ increases with decreasing Nb:STO layer thickness ($d$). As $d$ decreases to 1 unit cell (u.c.), $|S|$ reaches almost 500 $\mu$V/K. Although this systematic increase of $|S|$ is quite intriguing, its physical origin has not yet been elucidated.

In this paper, we investigate the electrodynamic responses of SLs using optical spectroscopy. As $d$ decreases, $m^*$ and $\tau$ are enhanced threefold due to a dimensional crossover of the polaron from 3D to quasi-2D. These changes in electrodynamic quantities result in a substantial enhancement of $|S|$, while $\mu$ remains constant. Our work suggests the potential importance of strong electron-phonon coupling in the search for highly efficient thermoelectric materials.

II. EXPERIMENTS

High-quality Nb:STO/STO SLs on single u.c. stepped $LaAlO_3$ (LAO) (001) substrates were fabricated using a pulsed laser deposition (PLD) technique at 950°C with in-situ monitoring of the specular spot intensity of reflection high-energy electron diffraction (RHEED). A LAO substrate was chosen for its simple and sharp optical phonon absorption lines which have little $T$-dependence. We investigated five different Nb:STO/STO SLs in this study. The thickness of the Nb:STO layer was systematically increased, to 1, 2, 3, 7, and 11 u.c. The number of insulating STO u.c. varied from 7 to 15, which is sufficiently thick to prevent any coupling between the doped layers. The repetition of each layer in the SLs was fixed at 20. Topographic images and x-ray diffraction results showed the high quality of the interfaces and surfaces in our SLs. For example, an atomic force microscope topographic image for (Nb:STO)1/(STO)15 is shown in Fig. 1(a). This image shows that the single u.c. stepped terraces were conserved after the deposition of the superlattices. Figure 1(b) shows an out-of-plane x-ray $\theta$-$2\theta$ diffraction pattern for the same superlattice. The satellite peaks are readily apparent, indicating the high quality of the interfaces and surfaces in our superlattices. The weak satellite peaks are probably due to small chemical contrast between the doped STO layer and the STO layer. Further details on the characterization of similar superlattices can be found elsewhere.[10, 11]

Near-normal incident in-plane reflectance spectra ($R(\omega)$) were measured between 3.7 meV and 6.9 eV using Fourier transform spectrophotometers and a grating monochromator. The symbols in Fig. 2 show the experimentally measured $R(\omega)$ of the SLs at room $T$. Due to the large penetration depth of light in the far-infrared region, the experimental $R(\omega)$ spectra contain features of three perovskite LAO substrate phonons. Since the wavelength of the light was much longer than the thicknesses of the SL layers, the light could not probe the individual layers. Instead, the whole SL layers should act as a homogeneous medium whose response is an average of those from individual layers.[12] Therefore, we could assume that the SLs would effectively behave as a single layer on the LAO substrate. We calculated $R(\omega)$ by further assuming that the in-plane optical conductivity spectra, $\sigma_1(\omega)$, of the effective medium could be represented by a Drude and several Lorentz oscillators:

$$\sigma_1(\omega) = \frac{e^2}{m^*}\frac{n_D\gamma_D}{\omega^2+\gamma_D^2} + \frac{e^2}{m^*}\sum_j \frac{n_j\gamma_j\omega^2}{\left(\omega_j^2-\omega^2\right)^2+\gamma_j^2\omega^2}, \qquad (2)$$

where $\gamma_j$ and $\omega_j$ are the scattering rate and resonant frequency of the $j$-th oscillator, respectively. The first term describes the coherent response from free carriers, while the second term describes the incoherent response due to optical phonons and bound charges. The lines in Fig. 2 show the curve-fitted results for $R(\omega)$. As shown in this figure, our analysis, based on the Drude-Lorentz model, provided a good description of the experimental $R(\omega)$ results for the Nb:STO/STO SLs. To fit $R(\omega)$ reasonably well up to 3.2 eV (the fundamental gap), we needed five oscillators including the Drude term, 3 phonons, and one broad mid-infrared (MIR) absorption peak.

## III. RESULTS AND DISCUSSION

Figure 3 illustrates $\sigma_1(\omega)$ of our Nb:STO/STO SLs. Each $\sigma_1(\omega)$ features 3 sharp perovskite phonon peaks, located near 15, 22, and 68 meV. The remaining broad spectral features can be decomposed into two peaks: *i.e.*, a Drude peak and a broad MIR peak located between ~ 0.15 and ~ 0.30 eV. As an example, we decompose the $\sigma_1(\omega)$ of the SL with 1 u.c. Nb:STO layers. The thin and thick grey lines, displayed at the bottom of Fig. 3, represent the Drude and MIR peaks, respectively. Note that the observed MIR peak structures are quite similar to those observed with other doped STOs, particularly in bulk systems,[13-15] where they have been typically attributed to a multi-phonon absorption peak due to the polaron.[14-16]

A polaron is a quasi-particle that can be described as an electron surrounded by a cloud of phonons.[17] The strong electron-phonon coupling of the polaron induces changes in the electrodynamic and optical properties. One of its most important changes is an increased inert mass, $m^*$. For an electron under a strong coupling, movement is impeded, so $m^*$ will increase. Therefore the strong electron-phonon coupling renormalizes the coherent Drude part of the optical spectra by $m/m^*$, whose spectral weight moves to higher energies as multi-phonon absorption bands in the MIR region.[17] In addition, collisions with other scattering centers will be less likely to occur, and $\tau$ will become larger. Therefore, according to polaron electrodynamics the change of $\tau$ should have close relationship with the mass enhancement.[18, 19]

As mentioned before, the optical response of the whole SL layers should act as an average of those from individual layers in the long wavelength limit. Therefore, the observed spectra in Fig. 3 might just come from the average of bulk responses of Nb:STO and STO layers. To check this possibility, we simulated the experimental $\sigma_1(\omega)$ using a multilayer simulation for the SL with one u.c. Nb:STO layer. We used a 2D effective medium approximation (2D EMA):[20]

$$\tilde{\varepsilon}_{SL}^{eff} = \frac{\tilde{\varepsilon}_{STO} d_{STO} + \tilde{\varepsilon}_{Nb:STO} d_{Nb:STO}}{d_{STO} + d_{Nb:STO}}, \qquad (3)$$

where $\tilde{\varepsilon}$ and $d$ correspond to the in-plane complex dielectric constants ($\tilde{\varepsilon} = \varepsilon_1 + i\varepsilon_2$, $\sigma_1(\omega) = \omega\varepsilon_2$) and the thickness of each layer, respectively. We used the dielectric constants of Nb:STO and STO obtained by optically measuring thick single-phase films separately deposited on LAO substrates. As shown in Fig. 4, it was quite evident that the $\sigma_1(\omega)$ from the 2D EMA could not reproduce the experimental $\sigma_1(\omega)$, especially in the MIR region. This disagreement suggests that the electrodynamics of the charge carriers inside the superlattice should undergo a substantial modification as the conducting Nb:STO layer becomes thin, *i.e.*, when the 2D character in the system becomes more dominant.

Note that the $\sigma_1(\omega)$ of STO is featureless up to 3.2 eV, except for the phonon modes below 0.1 eV, as shown in Fig. 4. Therefore, the spectral features, i.e. the Drude and MIR peaks in Fig. 3, should mainly arise from the $\sigma_1(\omega)$ of the Nb:STO layer and not from the STO layer. For this reason, the significant modification of the charge dynamics in the SL should be associated with the modification of the Nb:STO layer properties.

We calculated $m^*/m$ for our Nb:STO/STO SLs from $SW_{\mathrm{Drude}}$ and $SW_{\mathrm{MIR}}$, which are the spectral weights ($SW \equiv \int \sigma_1(\omega)d\omega$) of the Drude and MIR absorption peaks, respectively. With increasing coupling strength, both $SW$ redistributions, and hence $m^*/m$, becomes larger. Based on the experimental $\sigma_1(\omega)$, we evaluated $SW_{\mathrm{Drude}}$ and $SW_{\mathrm{MIR}}$. Figure 5 shows how the total electronic spectral weight ($SW_{\mathrm{total}} = SW_{\mathrm{Drude}} + SW_{\mathrm{MIR}}$) is almost linearly proportional to the volume fraction of the Nb:STO layers, satisfying the optical sum rule. By contrast, $SW_{\mathrm{Drude}}$ has a much stronger dependence on the volume fraction, while $SW_{\mathrm{MIR}}$ has a weaker dependence. The value for the polaron mass enhancement was obtained using the relationship $m^*/m = SW_{\mathrm{total}} / SW_{\mathrm{Drude}}$ (Ref. 14). Figure 6(a) shows the values of $m^*/m$ for our Nb:STO/STO SLs. As $d$ decreases from 11 to 1 u.c. within the SL, the $m^*/m$ value increases almost threefold, *i.e.*, from $2.0 \pm 0.8$ to $5.9 \pm 0.4$.

We can also determine $\tau$ from $\sigma_1(\omega)$ of the Drude peak. Note that the scattering rate $1/\tau$ of the free carrier corresponds to the half width at the half maximum of the Drude peak. The filled triangles in Fig. 6(b) represents the experimentally determined values of $\tau$. As $d$ decreases from 11 to 1 u.c., $\tau$ also increases almost threefold; in other words, $\tau$ is approximately proportional to $m^*/m$. This striking behavior is opposite to what would be expected from the electrodynamics of carriers in a typical metallic film: with decrease of $d$, $\tau$ should decrease due to enhanced surface/interface scattering. The filled triangles in Fig. 6(b) show the predictions of Matthiessen's rule, ($1/\tau = 1/\tau_{bulk} + v_F/d$, where $v_F$ is the Fermi velocity),[21] which takes into account the surface scattering in an ordinary metal. The increase of $\tau$ in our SLs cannot be explained by such a simple scattering mechanism involving free carriers. On the other hand, our observation of $\tau \propto m^*/m$ is consistent with theoretical predictions from polaron electrodynamics.[18, 19]

The substantial enhancement of $m^*/m$ and $\tau$ in our Nb:STO/STO SLs should be originated from a dimensional crossover of the polaron from 3D to quasi-2D. Specifically, with decreasing $d$, polarons in the Nb:STO layer should be more restricted in moving around due to the confined geometry of the conducting layer. The dimensional effects on polaron electrodynamics have not yet been reported experimentally. However, it has been theoretically predicted that $m^*/m$ should be enhanced as the dimensionality of the system is reduced. According to Feynmann and Peeters *et al.*, $m^*/m$ for the 3D and 2D cases exhibits an increase with the polaron coupling constant $\alpha$, as shown by the solid and dashed lines in the inset of Fig. 6(a), respectively.[22, 23] When $\alpha > \sim 1.2$ the corresponding $m^*/m$ value in 2D can be much larger than in 3D. Thus, $m^*/m$ should increase with decreasing $d$ in Nb:STO/STO SLs.

The enhanced $m^*$ due to the dimensional crossover of the polaron plays an important role in enhancing $|S|$ in Nb:STO/STO SLs. As shown in Fig. 6(a), the measured values of $|S|$ increase with decreasing $d$, which is quite similar to the $d$-dependence of $m^*/m$. Such similarity indicates that mass enhancement due to electron-phonon coupling could enhance $|S|$ (ref. 24).

The correlation between the $T$-dependent $m^*/m$ and $|S|$ values further supports our hypothesis, as shown in Fig. 7. The inset shows the $T$-dependent $|S|$ behavior for the SLs. The SL with 1 u.c. Nb:STO layers shows a distinctive peak at ~170 K, which is caused either by the strong phonon-drag effect or structural transition. Such $T$-dependent $|S|$ behavior has also been previously reported for an SL with one u.c. of Nb:STO, with a peak at ~150 K.[10] The $T$-dependent $m^*/m$ was able to reproduce the $T$-dependent behavior of $|S|$ in a qualitative level, also showing a peak at ~150 K. On the other hand, The SL with 11 u.c. Nb:STO layers do not exhibit any distinctive anomaly, also consistent with the $T$-dependent behavior of $|S|$. This result confirms that the effect of electron-phonon coupling can contribute to $|S|$, where an unexpectedly large $m^*/m$ can give rise to a large value of $|S|$.

The mechanism of enhancing $|S|$ through formation of polarons has another important aspect which cannot be obtained in the conventional approach to enhance $m^*$ by confined geometry or correlated electrons. As shown in Fig. 6(c), $\mu$ of the system remains nearly constant although $m^*$ is increased. This is because $\tau$ becomes renormalized along with $m^*$ in polaron electrodynamics. Thus $\sigma_{dc}$ should be linearly proportional to $n$ only, and we can obtain an enhancement of $S$ value without decreasing $\sigma_{dc}$ in the $zT$ value, *i.e.* Eq. (1). To date, efforts to enhance the efficiency of thermoelectric materials have mostly focused on engineering either the DOS near the $E_F$ in semiconductors or the strong electron correlation in oxides. However, using such conventional approaches, it would be difficult to enhance $S$ without affecting $\mu$ in $zT$. Our work suggests that $S$ can be enhanced without affecting $\mu$ when we use electron-phonon coupling to enhance $m^*$. We would argue that materials with strong electron-phonon coupling could be good candidates in the search for more efficient thermoelectric materials.

## IV. SUMMARY

We investigated the polaron dynamics of thermoelectric Nb:$SrTiO_3$/$SrTiO_3$ superlattices using optical spectroscopy. As the thickness of the Nb:$SrTiO_3$ layer decreased, we observed a dimensional crossover of polarons from 3D to quasi-2D. Due to the dimensional crossover, the polaron effective mass increased along with the relaxation time. The increased polaron effective mass seems to be the origin of the large enhancement of the Seebeck coefficient in the oxide superlattice. Our work suggests that materials with strong electron-phonon coupling could be good candidates in the search for more efficient thermoelectric materials.

## ACKNOWLEDGEMENTS

The authors are grateful for valuable discussions with A. Bratkovsky, S. Mukerjee, C. Bernhard, S. S. A. Seo, H.-Y. Choi, and S. W. Kim. This research was supported by the Basic Science Research Program through the National Research Foundation of Korea (NRF), funded by the Ministry of Education, Science and Technology (MEST) (No. 2009-0080567). Y.S.L. was supported by the Soongsil University Research Fund and H.O. was supported by Ministry of Education, Culture, Sports, Science and Technology (MEXT) (No. 20047007). The experiments at Pohang Accelerator Laboratory were supported in part by MEST and Pohang University of Science and Technology.

## Figure Captions

FIG. 1. (Color online) (a) Atomic force microscope topographic image and (b) x-ray $\theta$-$2\theta$ diffraction pattern for (Nb:STO)1/(STO)15.

FIG. 2. (Color online) $R(\omega)$ for Nb:STO/STO superlattices. The experimental $R(\omega)$ and Drude-Lorentz oscillator fit $R(\omega)$ spectra for different superlattices are shown, represented by different symbols and lines, respectively. For comparison, the $R(\omega)$ spectrum for a bare LAO substrate is indicated as a grey line.

FIG. 3. (Color online) $\sigma_1(\omega)$ for the Nb:STO/STO SLs. For the SL with 1 u.c. Nb:STO layers, the Drude and MIR peaks are represented by the thin and thick grey lines, respectively.

FIG. 4. (Color online) $\sigma_1(\omega)$ for the superlattices and thin films. The $\sigma_1(\omega)$ spectra of Nb:STO and STO single layers deposited on a LAO substrate are represented by a grey dashed line and a black dotted line, respectively. A macroscopic multilayer simulation using the $\sigma_1(\omega)$ spectrum of the Nb:STO and STO layers stacked vertically at ratios of 1 to 15 is shown by a thick solid cyan line. The experimental $\sigma_1(\omega)$ of the (Nb:STO)1/(STO)15 superlattice is shown by a thin solid red line.

FIG. 5. Electronic *SW* for the Nb:STO/STO SLs as a function of volume fraction for Nb:STO layers. The filled grey squares and empty light grey squares indicate $SW_{\text{Drude}}$ and $SW_{\text{MIR}}$, respectively. $SW_{\text{total}}$ (= $SW_{\text{Drude}} + SW_{\text{MIR}}$) is indicated by the filled black squares.

FIG. 6. (Color online) (a) $m^*/m$ (filled black circles) and $|S|$ (empty circles) for the Nb:STO/STO SLs as functions of the number of Nb:STO unit cells. The inset shows the theoretically calculated $m^*/m$ for the 3D (thick black line) and 2D (thin red line) cases as functions of the polaron coupling constant $\alpha$ (Ref. 22, 23). (b) Experimental $\tau$ values for the polaronic carriers (filled black triangles) and calculated $\tau$ values from Matthiessen's rule (empty grey triangles). The horizontal black dashed line represents the bulk limit, while the grey line is a guideline. **(c)** $\mu$ values of Nb:STO SLs.

FIG. 7. (Color online) *T*-dependent $m^*/m$ for SLs with 1 u.c. (red circles) and 11 u.c. (purple diamonds) of Nb:STO layers. The inset shows the *T*-dependent $|S|$ for the SLs. The arrow indicates the peak in the *T*-dependent $|S|$.

[a)] Present address: Department of Physics, University of California, San Diego, La Jolla, CA 92093, USA
[*] Electronic address: twnoh@snu.ac.kr

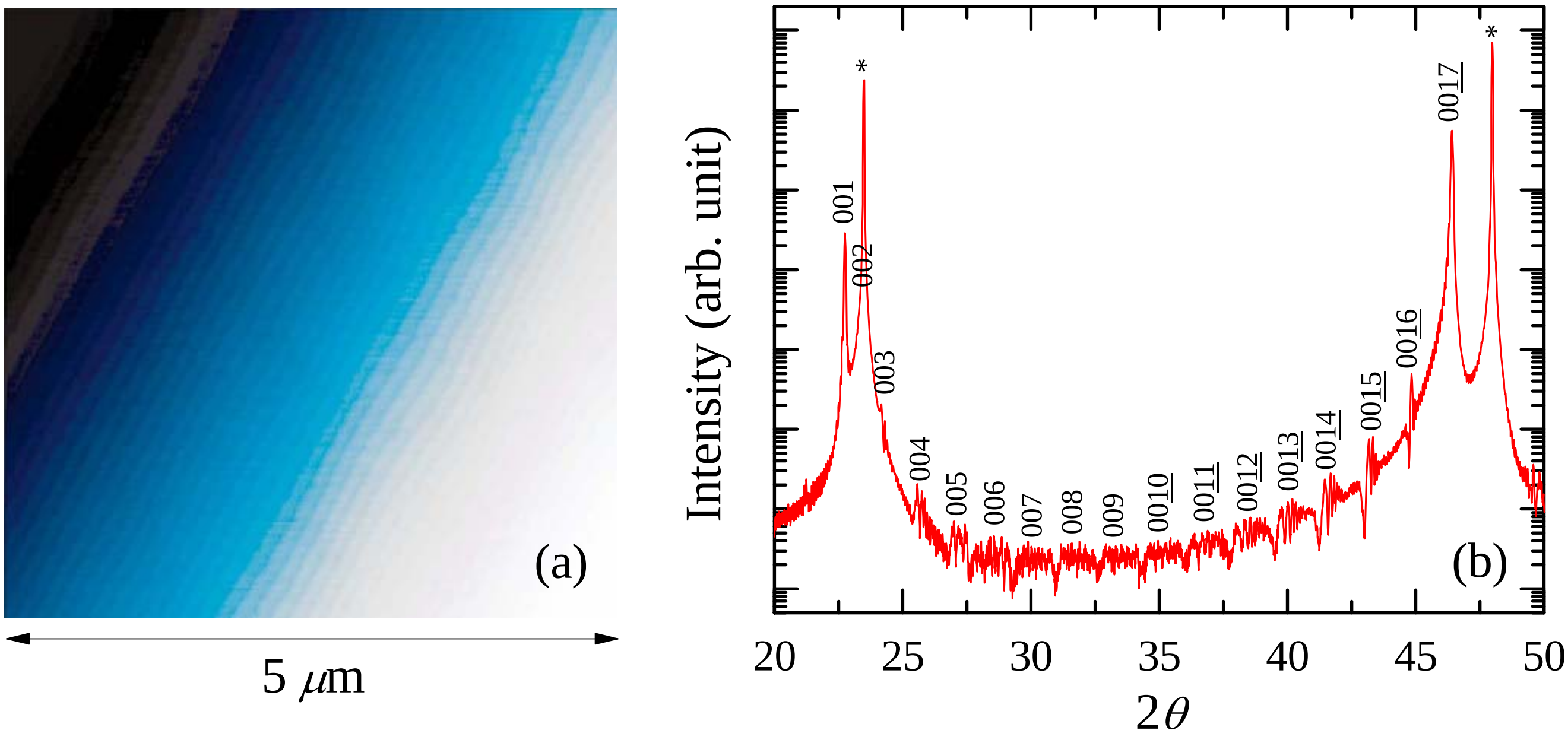

Fig. 1
Choi *et al.*

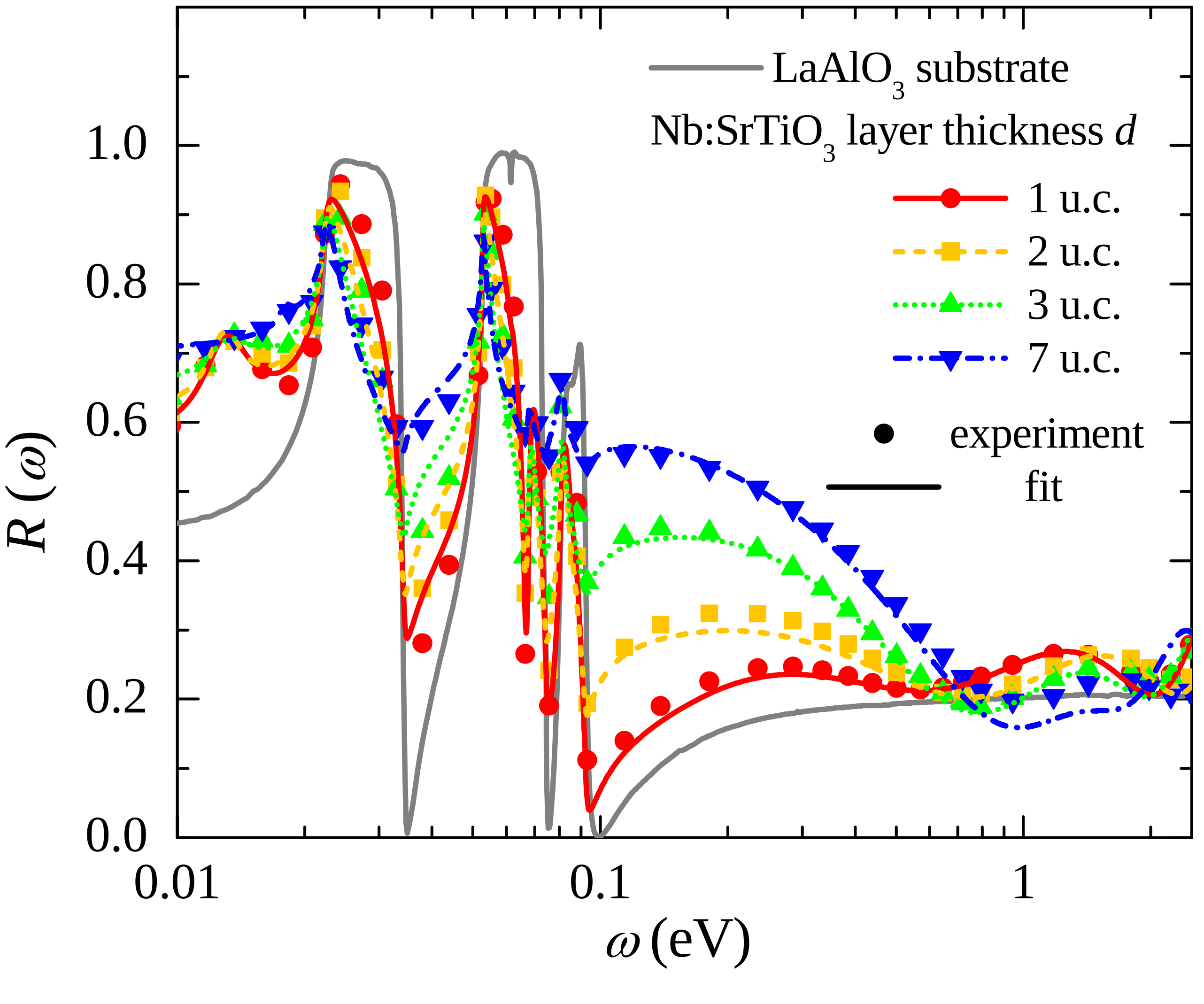

Fig. 2
Choi *et al.*

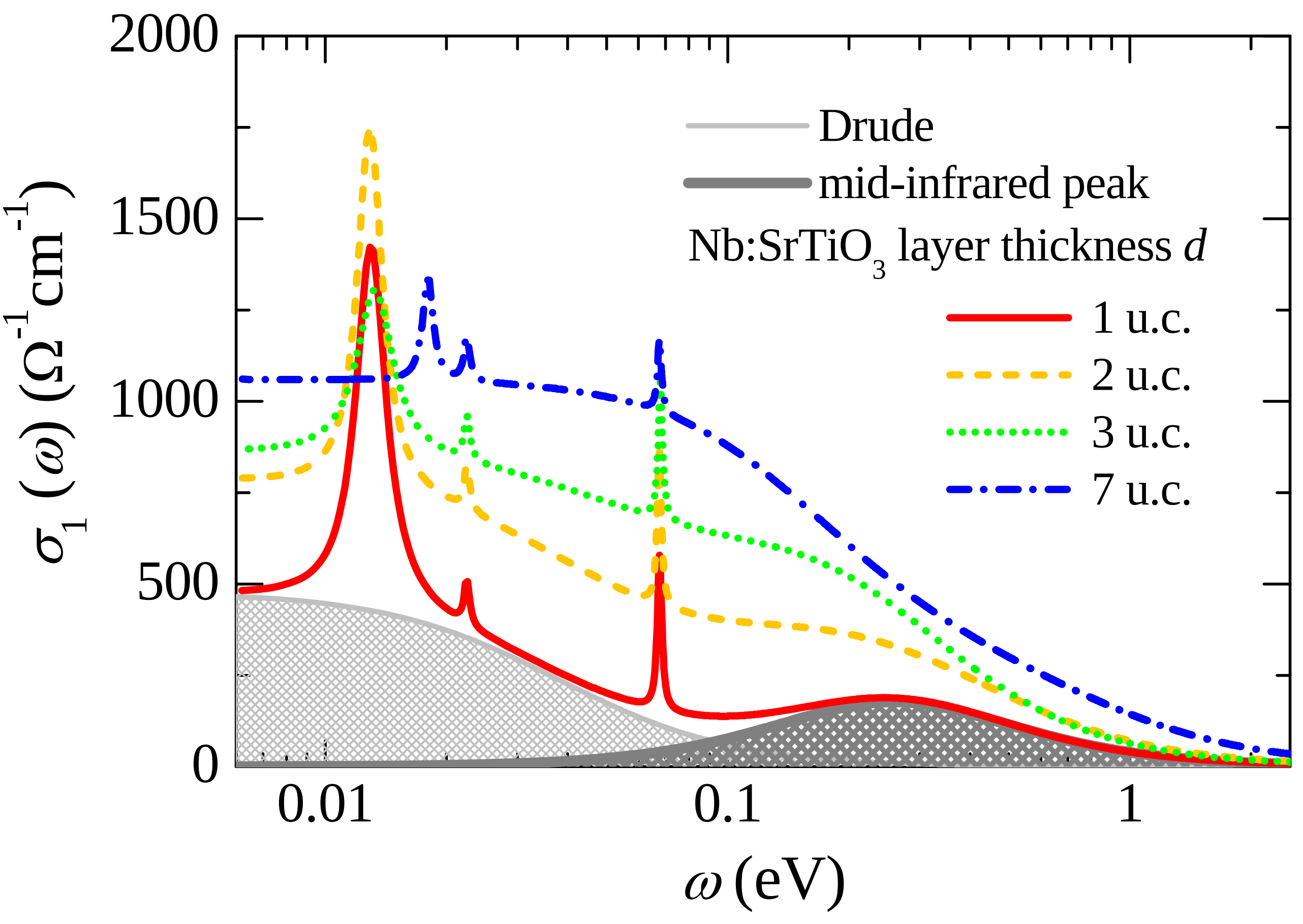

Fig. 3
Choi *et al.*

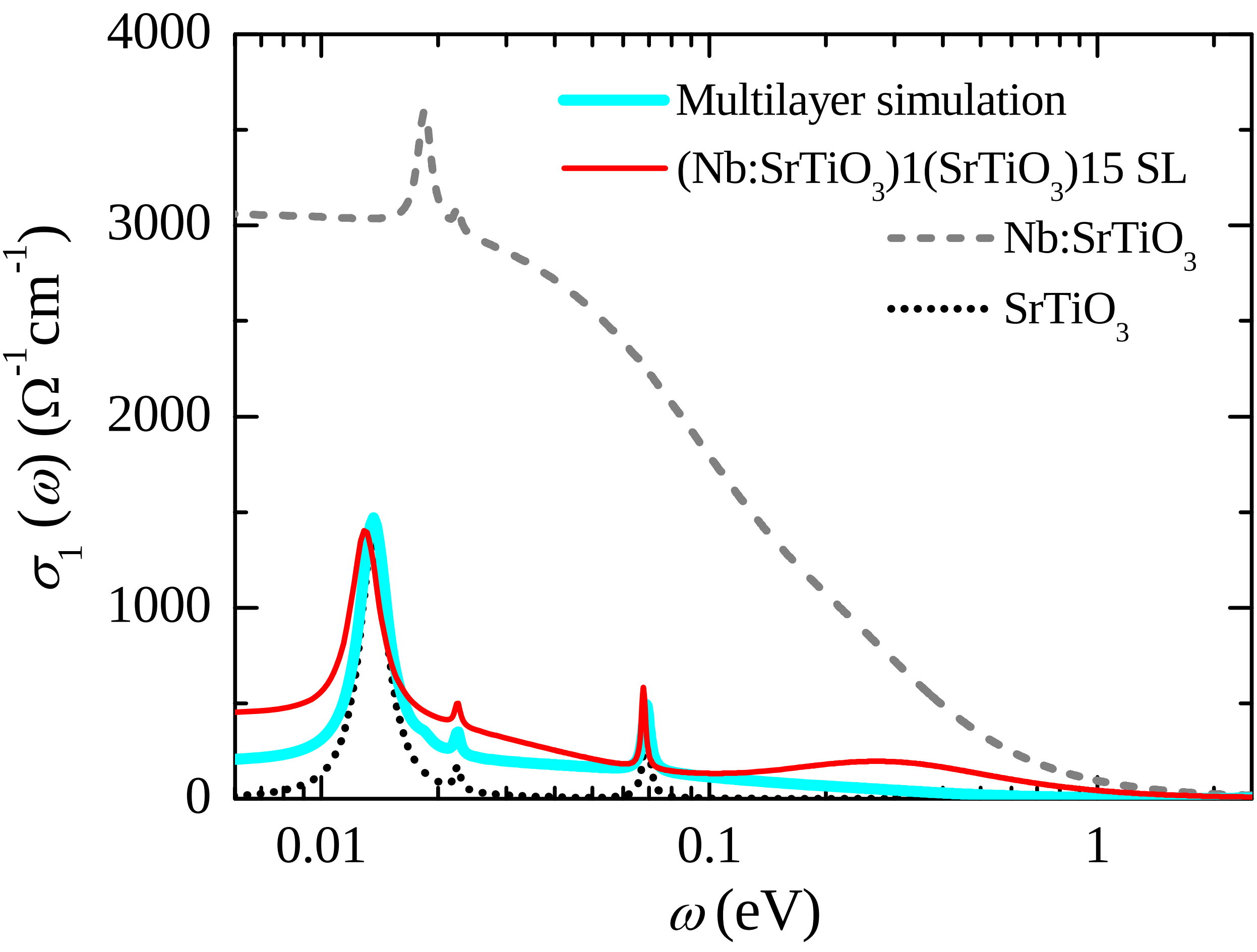

Fig. 4
Choi *et al.*

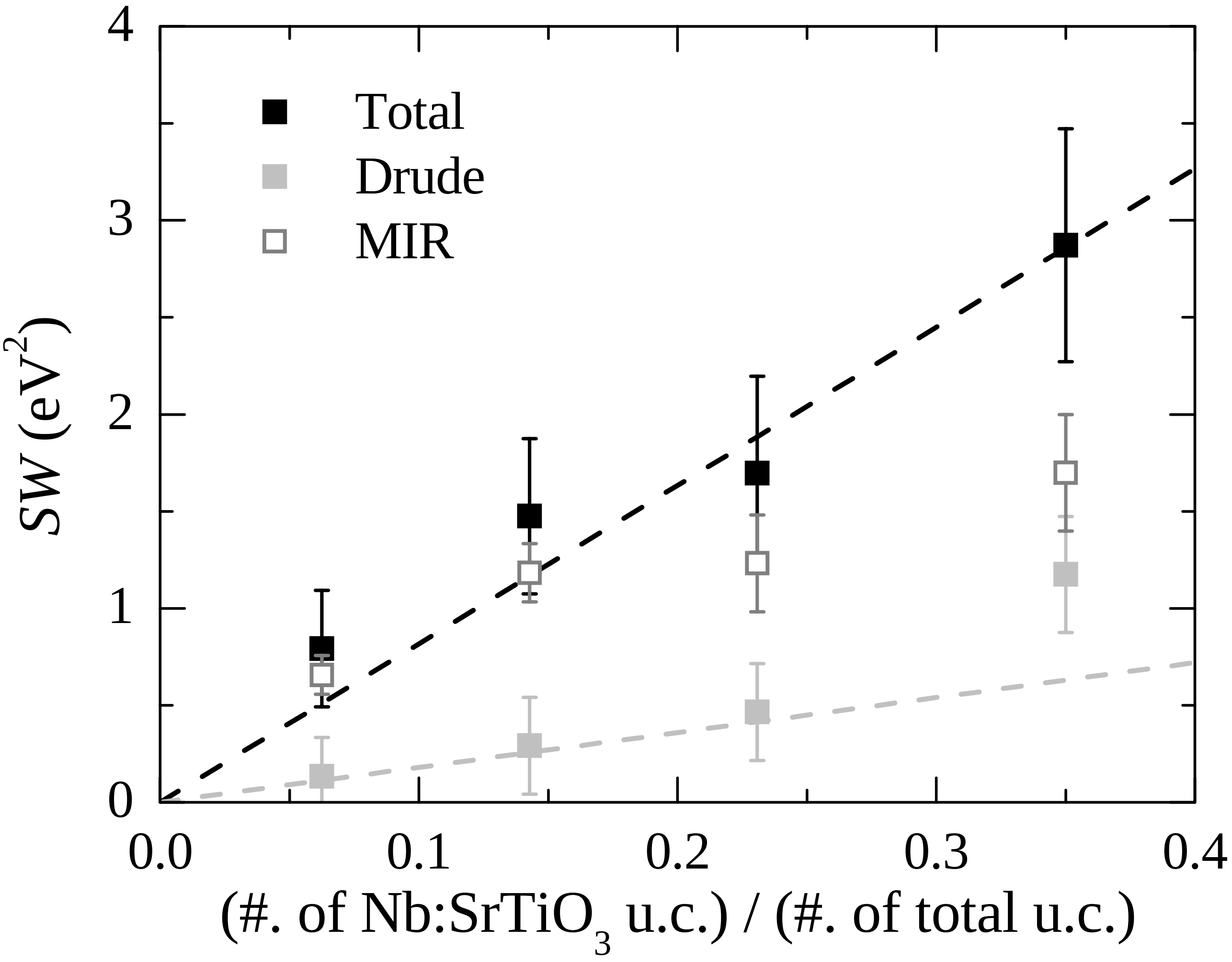

Fig. 5
Choi *et al.*

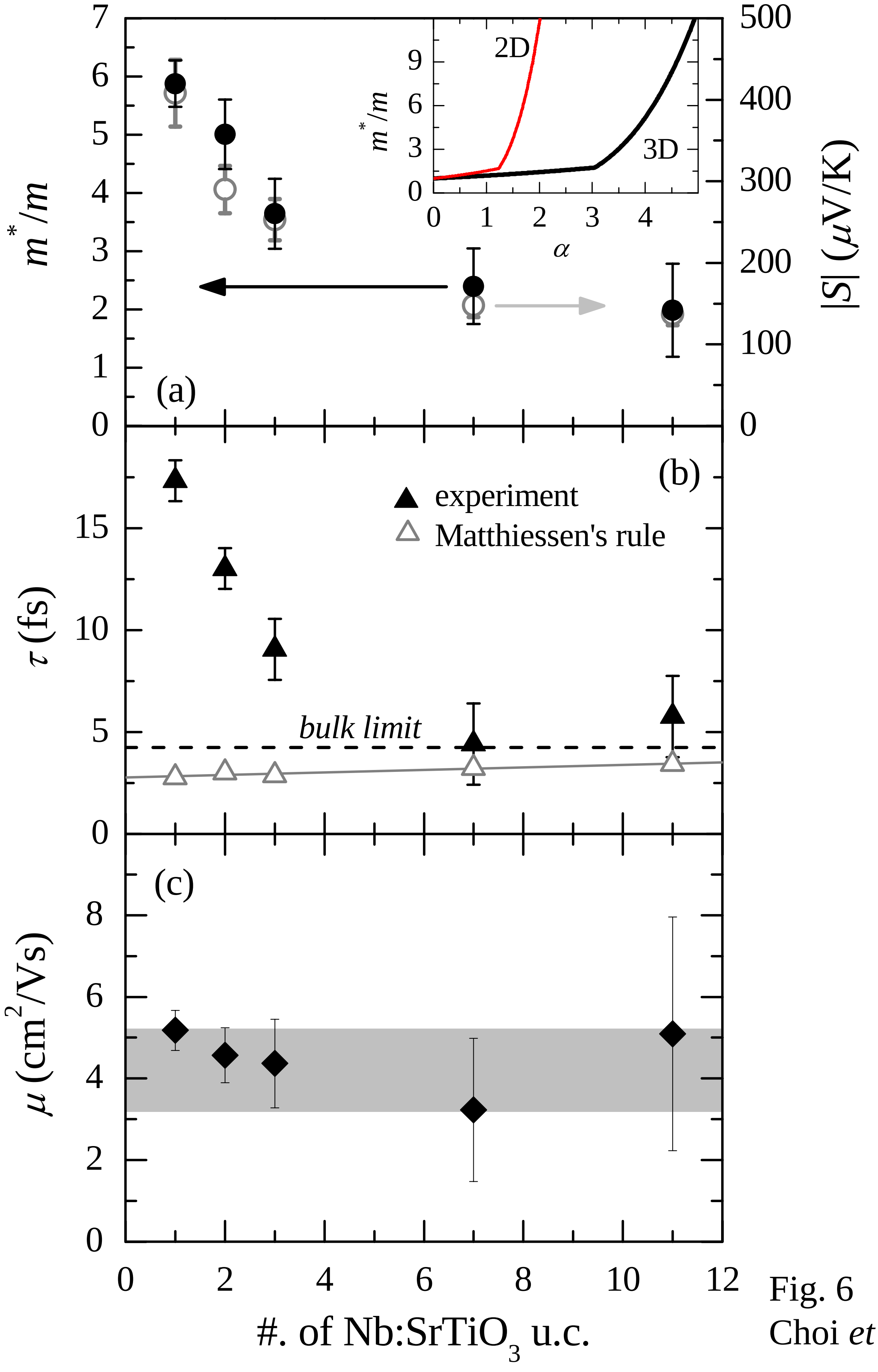

Fig. 6
Choi *et al.*

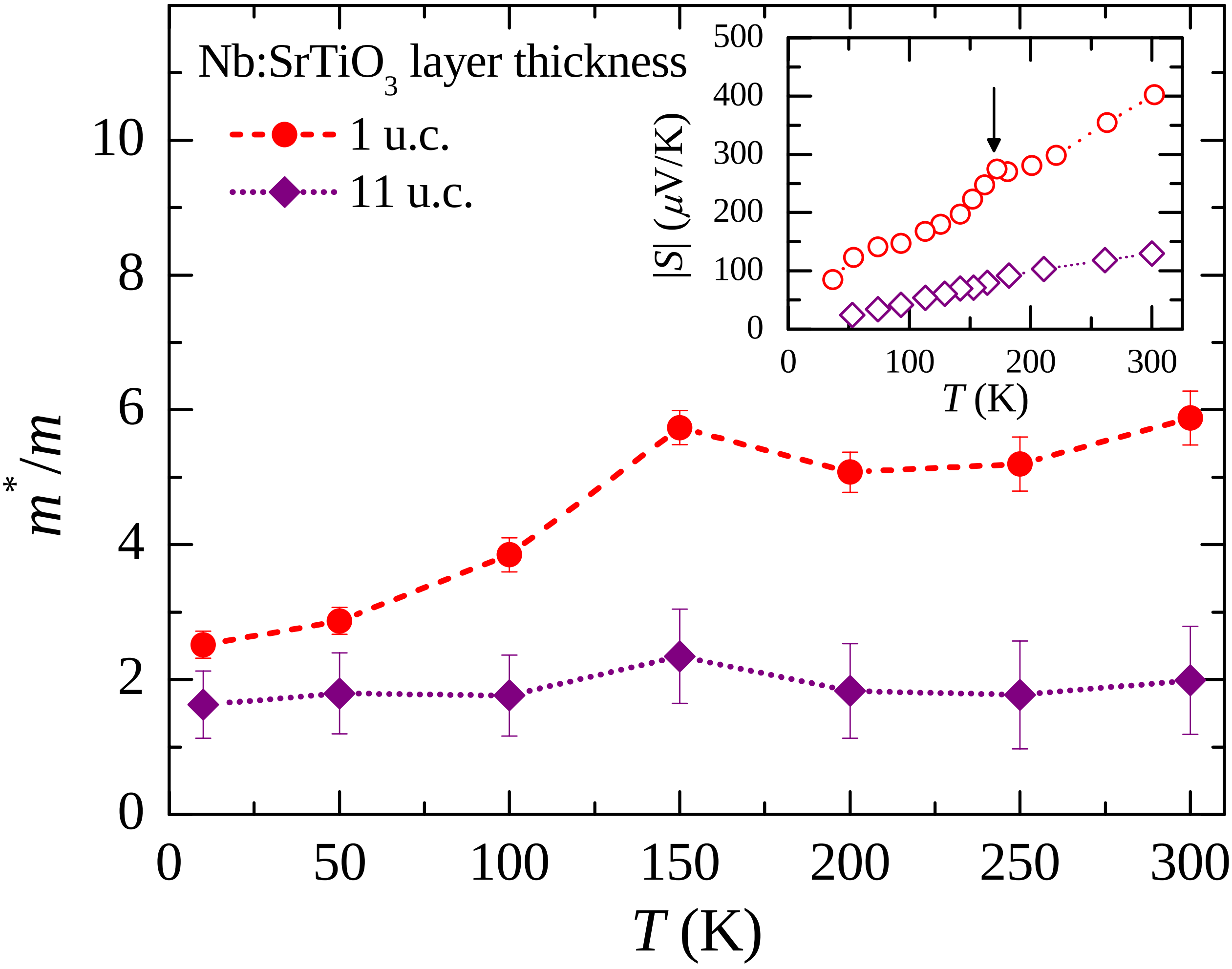

Fig. 7
Choi *et al.*